\documentclass[12pt]{iopart}
\expandafter\let\csname equation*\endcsname\relax
\expandafter\let\csname endequation*\endcsname\relax
\usepackage{amsmath, amsfonts}
\usepackage{framed}
\usepackage[T1]{fontenc}
\usepackage{times} 
\usepackage[cp1250]{inputenc}  
\advance\topmargin by -\headheight
\advance\topmargin by -\headsep
\evensidemargin=\oddsidemargin
\newcommand{\CA}{\mathcal{A}} 
\newcommand{\CH}{\mathcal{H}} 
\newcommand{\Z}{\mathbb{Z}} 
\renewcommand{\H}{\mathcal{H}} 
\newcommand{\oh}{{\tfrac{1}{2}}} 
\newcommand{\C}{\mathbb{C}} 
\newcommand{\sq}{\unskip\nobreak\kern5pt\nobreak\vrule height4pt width4pt depth0pt} 
\newcommand{\id}{\hbox{id}}
\newbox\ncintdbox \newbox\ncinttbox
\setbox0=\hbox{$-$} \setbox2=\hbox{$\displaystyle\int$}
\setbox\ncintdbox=\hbox{\rlap{\hbox
    to \wd2{\kern-.1em\box2\relax\hfil}}\box0\kern.1em}
\setbox0=\hbox{$\vcenter{\hrule width 4pt}$}
\setbox2=\hbox{$\textstyle\int$}
\setbox\ncinttbox=\hbox{\rlap{\hbox
    to \wd2{\kern-.14em\box2\relax\hfil}}\box0\kern.1em}

\begin{document}
\title{On almost commutative Friedmann–Lemaître–Robertson–Walker geometries.}
\author{Andrzej Sitarz\footnote{Partially supported by Polish National Science Center (NCN) grant 2016/21/B/ST1/02438}}
\address{Institute of Physics, Jagiellonian University,
	prof.\ Stanis\l awa \L ojasiewicza 11,\\ 30-348 Krak\'ow, Poland, \\ 
	Institute of Mathematics of the Polish Academy of Sciences, \'Sniadeckich 8, \\ 00-656 Warszawa, Poland.}
\begin{abstract} 
We analyze the leading terms of the spectral action for a model of noncommutative geometry, which is a product of $4$-dimensional Riemannian manifold with a two-point
space exploring the previously neglected case when the metrics over each sheet are 
different. Assuming the Friedmann–Lemaître–Robertson–Walker type of the metric for both sheets we obtain the action, which in addition to the the usual cosmological constant terms and the Einstein-Hilbert term involves a nonlinear interaction term. We study qualitative
picture of potential consequences of such term in the basic cosmological models.
\end{abstract}

\section{Introduction}

Cosmological models are based on Einstein equations, which link the
geometry of the universe with the energy-momentum density containing
the matter, radiation and dark energy (cosmological constant) contribution.
Such models have been thoroughly studied for both the standard equations
originating from the Einstein-Hilbert action as well as possible modifications 
of gravity, yet all of them are based rather on classical extensions of 
spacetime geometry then on modifying the basic formulations of 
geometry.

Noncommutative geometry \cite{Co95,Co96, CM08}, which has been studied extensively in the
physical context rather in relation to fundamental interactions of elementary
particles, offers a new insight into our understanding of the metric. In particular
some of the simplest models are of Kaluza-Klein type, with the extra dimensions
being of the finite type (that is consisting of finite number of points). This 
allows to study some new effects and effectively draw some basic 
conclusions that could have cosmological implications.

In this paper we aim to study the simplest geometry of the product of
a spacetime with a two-point space. That roughly corresponds to the 
particle physics Connes-Lott model studied in the noncommutative 
setup \cite{CoLo}, where the two points are reflected in the chirality of the 
fundamental fermions. However, contrary to the usual assumptions 
we want to investigate metrics that are not of product type, that is
they might differ on the two sheets of spacetime. As the internal 
metric between the points has the natural interpretation of the Higgs 
field, we shall see that the natural generalization of the Einstein-Hilbert 
action introduces a new term that in the broken symmetry phase allows
for the interaction between the metrics.

The paper is organized as follows: first we present the basic tools and
notation for the geometry studied including the spectral triple of the
model and the spectral action. We present the effective methods of 
computing the action using the Wodzicki residue over the 
pseudodifferential calculus of symbols and derive the action functional 
for the model of  Friedmann–Lemaître–Robertson–Walker type geometries. Finally we study 
the equations of motions and analyze few cases of cosmological models.

\section{Almost commutative geometries and spectral action}

An almost commutative geometry is a model based on the product geometry 
of the compact Riemannian spin manifold with a finite dimensional space 
(not necesarily commutative) which is described throug a finite-dimensional 
spectral triple.  Such model was among the first ones to be considered \cite{CoLo} 
by Connes-Lott and has led to the interpretation of 
the Higgs field as a connection arising from the geometry of the finite space.
That is the minimal noncommutative extension of the classical geometry,
which is basically of Kaluza-Klein type, however, with the internal space
that is not a manifold. The simplest version of the discrete geometry is a 
two-point space described by its algebra  of complex-valued functions 
$A_F = \C \oplus \C$.   

Although such ,,spaces'' are not described  by the usual differential geometry, 
the noncommutative geometry offers a way to treat both manifolds 
as well discrete spaces and finite-dimensional algebras (not necessarily commutative) on equal footing.  Such noncommutative extension of the 
standard differential geometry uses the construction of spectral triples
\cite{CM08}. In short, a spectral triple $(A, \CH, D)$ consists of the following data:  
the algebra $A$ (which in the classical case is the algebra of smooth functions 
over the manifold), faithfully represented as bounded operators on the 
Hilbert space $\CH$,  and an unbounded selfadjoint operator, such that 
for every $a\! \in\! A$ the commutator $[D,\pi(a)]$ is bounded, where 
$\pi(a)$ denotes the representation.  The classical example of a spectral 
triple is provided by a compact Riemannian spin manifold $M$ and 
$(C^\infty(M), L^2(S), D)$, where  $L^2(S)$ denotes the Hilbert space 
of square-summable sections of spinors and $D$ is the usual Dirac 
operator.

The metric is then implicitly encoded in the Dirac operator $D$ and the gravity 
action functional is constructed from the spectral data of the Dirac operator,
for example, using the terms from the heat kernel asymptotic expansion of
 the operator $e^{-t D^2}$ \cite{Ka}.
 
\subsection{Spectral triples for almost commutative geometries} \label{s21}

We begin with a short presentation of a spectral triple that is a minimal noncommutative extension of the classical geometry and be the basis 
to study the models with a generalized Friedmann–Lemaître–Robertson–Walker type metric. 
Consider the algebra $\CA = C^\infty(M) \otimes (\C \oplus \C)$, represented 
on $L^2(S) \otimes \C^2$. The algebra can be seen an algebra of smooth  
functions on $M$ (which we assume to be even-dimensional) valued in 
the diagonal $2$ by $2$ complex matrices and its representation is then 
natural multiplication from the left on two copies 
of the spinor fields. As the geometry is in fact a product geometry and the 
underlying space is nothin but a Cartesian product  $M \times \Z_2$, where 
$\Z_2$ denotes the two-point space.

The usual product-type Dirac operator is taken as,
\begin{equation}
D_o = D \otimes \id +\, \gamma \otimes D_F, \label{Dipr}
\end{equation}
where $D_F$ the Dirac operator on the two-point space,
$$ 
D_F =  \left( \begin{array}{cc} 0 & \Phi \\ \Phi^* & 0 \end{array} \right) 
$$
with $\Phi \in C^\infty(M)$ understood as a complex scalar field that is 
identified with the Higgs field in this toy model.

However, it is easy to see that that the Dirac operator $D_o$ is not the
most general one, as the product metric is not the only metric that can
exist on the product of two metric spaces. To have a more general picture
let us consider now a slight modification of the product geometry  (\ref{Dipr}), 
allowing the full Dirac operator to be of the form:
\begin{equation}
D = \left( \begin{array}{cc} D_{1}  & \gamma \phi \\
\gamma \phi^* & D_{2}  \end{array} \right), 
\label{dirt}
\end{equation}
where $D_1,D_2$ are two independent Dirac operators on the manifold $M$. 
This is, in fact, the most general Dirac operator on the product manifold that 
we can consider, which gives the usual spectral triple when restricted to each 
point of the finite-dimensional space and the finite spectral triple for the discrete 
space alone.

The Dirac operator $D$ (\ref{dirt}) introduces a new range of problems to the model,
both in the physical interpretation as well as in computations. Concerning the latter
we first encounter the situation that each of the fibres over the two-point space
of the Hilbert space of spinors should be considered with a different scalar 
product. To avoid this issue one should use the unitary equivalence of the Hilbert
space, then, however, complicating the form of the Dirac operators. From the
point of view of the interpretation we have a model with two metrics, which
resembles the bimetric gravity models${}^\ddag$\footnotetext{The author thanks 
	Marco de Cesare for turning his the attention to it.}
 (see \cite{HR12,SMS16} and references therein)  and it remains a question of
choice, which one is the background metric.

The last crucial technical difference is in the form of the square of the Dirac
operator, which is
$$ D^2 =  \left( \begin{array}{cc} D_{1}^2  + \phi \phi^*& 
	\gamma (D_1 \phi - \phi D_2)\\
	- \gamma (D_1 \phi^* - \phi^* D_2) 
	& D_{2}^2 + \phi \phi^* \end{array} \right), 
$$
and in the case $D_1=D_2$ differs from the "usual" Dirac operator only by
terms of order $0$. In the general case some new terms of the first order
arise, raising also the question whether the full Dirac operator is 
torsion-free even if $D_1$ and $D_2$ were.

\subsection{The Euclidean Friedmann–Lemaître–Robertson–Walker geometry}
	
In what follows we shall discuss the Euclidean version of the 
Roberston-Walker geometry over the minimal noncommutative 
generalisation, deriving the corresponding action through the spectral 
action principle. 

We concentrate first on the flat, toroidal geometry, using as as the background
the Hilbert space of spinors with respect to the equivariant metric. This allows
us to compute the spectral action agains the time-independent metric and
derive the equations of motions.

Let us recall (compare \cite{Si14}, Lemma 3.1) the basic result. $D_g$ be 
the usual Dirac operator on $M$ of dimension $d$ with the metric 
$g_{ab}$ and $\H_g$ be the Hilbert space of $L^2(S,g)$ (where the measure 
is taken with respect to the metric $g$).  Then, if $h_{ab}$ is another metric 
on $M$ and $ \hbox{det}(g) = \chi \hbox{det}(h)$ then the 
operator acting on $\H_h$,
$$ D_h = \chi^{\frac{1}{2}} \tilde{D} \chi^{-\frac{1}{2}}, $$ 
is unitarily equivalent to $D$.

It is easy to find the explicit unitary equivalence, let $U\!: \Psi \to \chi^{-\frac{1}{2}} \Psi$ be an isometry map between Hilbert spaces $U\!: \H_g \to \H_h$, which
means,
$$|| \Psi ||_{\H_g} = \int_M \sqrt{g} |\Psi|^2 =  
\int_M \sqrt{g} \chi | \chi^{-\frac{1}{2}} \psi |^2  = ||U \Psi ||_{{\H_h}}. $$
Consequently, the operator $D_h = U^{-1} D_g U$ is an operator on $\H_h$, 
which is (by construction) unitary equivalent to $D_g$. 

The above procedure allows us to map Dirac operators to operators on 
spinor Hilbert spaces where the scalar product is given by a different metric.
For the Friedmann–Lemaître–Robertson–Walker type geometry we shall be interested in the case 
where $d\!=\!4$  and the conformal factor $\chi = a(t)^{-\frac{3}{4}}$, so that 
the Riemann measure on the torus is just the measure of the flat torus.

\subsubsection{Toroidal geometry}

Consider the toroidal euclidean Friedmann–Lemaître–Robertson–Walker geometry, which
is just a $4$-dimensional torus, with the metric of the following form:
$$ ds^2 = (dt)^2 + a(t)^2 \left(  (dx_1)^2 + (dx_2)^2 + (dx_3)^2  \right). $$
The Dirac operator for the above metric reads 
$$ D = \gamma^0 \partial_t + \frac{1}{a(t)} \left( 
\gamma^1 \partial_1 + \gamma^2 \partial_2 + 
\gamma^3 \partial_3  \right) +  \gamma^0 \frac{3\dot{a}(t)}{2a(t)}, 
$$
where we use antihermitian $\gamma$ matrices, so that $D$ is hermitian 
on the sections of the spinor bundle where the inner product is computed
with respect to the above metric.

Using the method described above we find that the formula (in local coordinates)
for the unitarily equivalent Dirac operator, $D_f$, over the flat torus becomes:
$$ 
D_f = a(t)^{-\frac{3}{2}} D a(t)^\frac{3}{2} = 
\gamma^0 \partial_t + \frac{1}{a(t)} \left( \gamma^1 \partial_1 + \gamma^2 \partial_2 + 
\gamma^3 \partial_3  \right).
$$ 

The minimal noncommutative generalisation, which we discussed in section 
\ref{s21} will have the form:

\begin{equation}
D = \left( \begin{array}{cc} D_{1,t}  +  \frac{1}{a_1(t)} D_3 & \gamma \phi \\
\gamma \phi^* & D_{2,t} + \frac{1}{a_2(t)} D_3 \end{array} \right), 
\label{dirt}
\end{equation}
with possibly  different scaling factors $a_i(t)$. Here, $D_3$ denotes the
the fixed Dirac operator on the three torus $D_3 = \gamma^1 \partial_1 + \gamma^2 \partial_2 + \gamma^3 \partial_3$ but the expression can be easily extended to
the case of spherical geometry. Then $D_3$ should be the Dirac operator on
the sphere (taken for the invariant metric and fixed radius of the sphere) acting
on the respective Hilbert space of spinors over $S^3$.

\section{Spectral action.}

In this section we briefly describe the methods to compute the first two 
leading terms of the spectral action for the toroidal model (which could 
be easily extended to the more general geometries of Friedmann–Lemaître–Robertson–Walker
models). The spectral action \cite{EI19} is usually presented as the asymptotic
expansion in $\Lambda$ of the  trace of $f(D^2 / \Lambda^2)$ for a suitable
function $f$ (for example, a smooth approximation of the step function). 
Using Mellin transform and heat trace expansion, the leading terms can be
expressed using Gilkey-Seeley-de Witt coefficients. For the pseudodifferential
operator one can equivalently use the formulation of the spectral action
using the Wodzicki residue, where the first two leading terms are:
\begin{equation}
{\mathcal{S}}(D) = \lambda^4 \; \hbox{Wres} (D^{-4}) + c \lambda^2 \hbox{Wres} (D^{-2}),
\end{equation} 
with $\Lambda$ being the scaling factor, which we interpret as related to some
cutoff energy scale and $c$ is an arbitrary coefficient related to the exact form
of the cutoff function  (see \cite{EI19} for details).

If $D^2$ is a differential operator that could be split into homogeneous parts
of order $2,1$ and $0$ respectively, and the symbols of $D^2$ are $\sigma(D^2)=a_2+a_1+a_0$, with $a_k$ homogeneous of degree $k$, 
then using the algebra of the pseudodifferential calculus we can compute 
the symbols of its inverse,
\begin{equation}
\begin{aligned}
&b_0  = (a_2)^{-1}, \\
&b_1  = -  \left( b_0 a_1 +  \partial^\xi_k(b_0) \partial_k(a_2) \right) b_0, \\
&b_2  = - \left(  b_1 a_1 + b_0 a_0 +  \partial^\xi_k(b_0) \partial_k(a_1) 
+  \partial^\xi_k(b_1) \partial_k(a_2) + \oh \partial^\xi_k \partial^\xi_j (b_0) \partial_k \partial_j (a_2) \right) b_0, 
\end{aligned}
\end{equation} 
where $\partial^\xi_k$ denotes partial derivative with respect to coordinate
of the cotangent bundle, and $b_k$ is homogeneous of degree $-2-k$. 

Since the Wodzicki residue of a pseudodifferential is proportional to the integral 
over the cosphere bundle of the symbol of degree $-4$ (for a $4$-dimensional manifold) we obtain that the spectral action (the first two leading terms ) become:

\begin{equation}
{\mathcal{S}}(D) =  \int_M \int_{|\xi|=1} \left( \lambda^4 (b_0)^2
+ c \lambda^2 b_2 \right). \label{spac}
\end{equation} 

\subsection{Action for the toroidal Friedmann–Lemaître–Robertson–Walker}

We begin with the explicit computations of the spectral action for the 
assumed toroidal almost-commutative Friedmann–Lemaître–Robertson–Walker geometry.  	
Assume that the the underlying geometry is $S^1 \times T^3$, with
constant metric of equal lenght along all directions and that
the Dirac operator is as in (\ref{dirt}) with:
\begin{equation}
\begin{aligned}
& D_3 = \gamma^1 \partial_1 +\gamma^2 \partial_2  + \gamma^3 \partial_3, \\
& D_{n,t} = \gamma^0 \left( \partial_t + H_n(t) \right) , \;\; n=1,2,
\end{aligned}
\end{equation}
where $H_1(t), H_2(t)$ are some functions. Note that technically, we are always
considering not a {\em true} Dirac operator, but its unitarily equivalent counterpart
on a different Hilbert space. For simplicity we could write $D$ as
$$ \gamma^0 (\partial_t + H(t)) 
+ A(t) D_3
+ \gamma F(t), $$
where
$$ 
H(t) = \left( \begin{array}{cc} H_1(t) &  0 \\ 0 & H_2(t) \end{array} \right),
\qquad
A(t) = \left( \begin{array}{cc} \frac{1}{a_1(t)} &  0 \\ 0 & \frac{1}{a_2(t)} \end{array} \right),
\qquad
F(t) = \left( \begin{array}{cc} 0 &  \Phi(t) \\ \Phi^*(t) & 0 \end{array} \right).
$$
First, we compute $D^2$:
\begin{equation}
\begin{aligned}
D^2 = & \; (\partial_t)^2 + A(t)^{2} (D_3)^2 \\
& + 2 H(t) \partial_t + \gamma^0 (\partial_t A(t)) D_3 + [F(t), A(t)]  \gamma D_3 \\
& + \partial_t H(t) + H(t)^2 + F(t)^2  
-  \gamma \gamma^0 \left( \partial_tF(t) + [ H(t), F(t)] \right).
\end{aligned}
\end{equation}

We compute now the two leading order terms of the spectral action using
the methods of the pseudodifferential calculus as presented above.

Let us first write the symbols of the differential operator $D$, splitting it
into the components, which are homogeneous in $\xi$.

$$
\begin{aligned}
&a_2 =   \xi_0^2 + A(t)^2 \left( (\xi_1)^2 +(\xi_2)^2 + (\xi_3)^2 \right) \\
&a_1 =    i  \left( - 2 H(t) \xi_0 + \partial_t A(t)  \gamma^0  \left( \gamma^i \xi_i  \right) 
+ [F,A] \gamma (\gamma^i \xi_i) \right), \\
& a_0 =  - H(t)^2 - \partial_t H(t) + F(t)^2 - \gamma \gamma^0 \left( 
\partial_t F + [H, F] \right).
\end{aligned}
$$

The symbol of $(D_H)^{-2}$ reads:
$$ \sigma(D_H^{-2}) = b_0 + b_1 + b_2 + \cdots, $$
where the part $b_0$ (homogeneous of order $-2$ and $b_2$, homogeneous
of order $-4$ are (after taking the trace over the endomorphisms of the spinor
bundle they are acting upon and using the periodicity of the trace). 
For convenience, we skip the explicit dependence on coordinate $t$ and 
denote $\xi^2 = \xi_1^2 +\xi_2^2+\xi_3^2$.

\begin{equation}
\begin{aligned}
 b_0(\xi) \;=\;& (\xi_0^2 + A(t) \xi^2)^{-1}, \\
 b_2^n(\xi) \;=\; & \left( \,b_0 \,[A,F]\, b_0  FA b_0 + b_0 F A b_0 [A, F] b_0 \right) \xi^2 - b_0^2 F^2, \\
 b_2^c(\xi) \;=\; & 
   \dot{A} A H (4 b_0^2 \xi^2 - 24 b_0^4 \xi^2 \xi_0^2) \\
& + \dot{A}^2 ( 8 A^2 b_0^4 (\xi^2)^2 - 48 A^2 b_0^5 (\xi^2)^2 \xi_0^2
                    - b_0^3 \xi^2 + 8 b_0^4 \xi^2 \xi_0^2 ) \\
 & + \ddot{A} (-2 A b_0^3 \xi^2 + 8 A b_0^4 \xi^2 \xi_0^2) \\
 & + H^2 (b_0^2 - 4 b_0^3 \xi_0^2) +\dot{H} (b_0^2 - 4 b_0^3 \xi_0^2).
\end{aligned}
\end{equation}
where we have split the $b_2$ term into the diagonal (commutative) term
$b_2^c$ and the noncommutative term $b_2^n$. 

 We compute first the diagonal, term, as it will we just a sum of two
 independent entries:
 
 $$ \int_{|\xi|=1} b_2^c(\xi) = 
     \hbox{tr} \left(  \frac{\pi^2}{A^5} \left( 3 \dot{A}^2 - A \ddot{A} \right) \right), $$
 
 For the part, which is nonscalar we first need to compute the trace,
 which leads us to the following expression:
 
 $$ \int_{|\xi|=1} |\Phi(t)|^2 \left( \left( \frac{1}{a_1} - \frac{1}{a_2} \right)^2
  b_0(a_1) b_0(a_2) \left( b_0(a_1) + b_0(a_2) \right)
  - b_0(a_1)^2 - b_0(a_2)^2 \right), $$
 which after integration gives:
 $$ 2 \pi^2 |\Phi|^2 \frac{(a_1-a_2)^2}{(a_1+a_2)} \left( {a_1^2} 
 + a_1 a_2 + {a_2}^2 \right) $$
 
We can compare now the above result to the classical Einstein-Hilbert action for the Friedmann–Lemaître–Robertson–Walker metric. The kinetic term is exactly the same as
the scalar curvature (multiplied by the volume form), up to a multiplicative
constant,
$$ \sqrt{g} R(g) = 6 \left(-3 \left( { \dot{A}(t)^2 \over A(t)^5 }\right) 
+ {\ddot{A}(t) \over A(t)^4} \right). $$
though, of course, we have two such terms, independently for $a_1(t)$ and
$a_2(t)$. The difference is the potential term, which describes the coupling
between the metric and the $\Phi$ field. The latter is naturally interpreted
as the Higgs field and therefore we can investigate what happens to the 
scaling factors if the vacuum expectation value of $\Phi$ is nonzero.

So the total spectral action (restricted to two leading terms), expressed explicitly 
in terms of $a_1(t)$ and $a_2(t)$ is:

\begin{equation}
\begin{aligned}
{\cal S} = & 2\pi^2 \int dt \biggl( \lambda^4 \left( a_1(t)^3 + a_2(t)^3 \right) \\
& - \lambda^2 c \left( \dot{a}_1(t) ^2 a_1(t) + \dot{a}_2(t)^2 a_2(t) \right) \\
&- \lambda^2 c |\Phi|^2 \left(  a_1(t)^3 + a_2(t)^3 \right) \\
&+ \left. \lambda^2 c |\Phi|^2 \left(\frac{(a_1(t)-a_2(t))^2}{(a_1(t)+a_2(t))} \left( {a_1(t)}^2 
+ a_1(t) a_2(t) + {a_2(t)}^2 \right) \right) \right) + {\cal L}_m,
\end{aligned} 
\end{equation}
where ${\cal L}_m$ denotes other terms that could arise either from some higher-order corrections of the spectral action or matter terms. We have omitted a total derivative 
term $\frac{d}{dt} (\dot{a} a^2)$. 

Let us note that the first and the third terms are exactly the same, so we obtain
only corrections to the cosmological constant. In fact, the presence of these terms
alone (in the nonzero Higgs vacuum expectation value) motivates the mere existence
of the cosmological constant term. Finally let us point out that the action is, of course,
Euclidean and in order to proceed with physical analysis we need to perform Wick rotation to the Lorentzian signature. In our case that will lead only to the change of
the sign in the dynamical part of the action. 

In our considerations we have neglected all terms with $H(t)$, which is a potential
torsion term that we have incorporated into the Dirac operator. However, since
the only terms that it appears are ,,diagonal'', that is it appear separately for each
sheet in our model and does not mix the scaling functions $a_1$ with $a_2$ we
assume it to vanish, similarly like in the classical case. 

Putting it all together we finally obtain the physical effective action for the
toroidal Friedmann–Lemaître–Robertson–Walker geometry as,

 \begin{equation}
 \begin{aligned}
 {\cal S} = & \Lambda \left( a_1(t)^3 + a_2(t)^3 \right) +
                   6 \left( \dot{a}_1(t) ^2 a_1(t) + \dot{a}_2(t)^2 a_2(t) \right) \\
&+ \left. \alpha |\Phi|^2 \left(\frac{(a_1(t)-a_2(t))^2}{(a_1(t)+a_2(t))} \left( {a_1(t)}^2 
 + a_1(t) a_2(t) + {a_2(t)}^2 \right) \right) \right) + {\cal L}_m,
 \label{2ptaction}
 \end{aligned} 
 \end{equation}
where we have introduced for simplicity effective constants $\Lambda$ (cosmological constant) and $\alpha$ (strength of the potential). In the rest of the paper we shall
briefly analyze the consequences of the extra interaction term between the two scales
$a_1(t)$ and $a_2(t)$.

\section{The equations of motion}

The Friedman equations of motion could be easily derived as the Euler-Lagrange equations from the action (\ref{2ptaction}). First, consider the classical case, with no 
noncommutativity. To have the full set of equations we need to conveniently 
express the Lagrangian density, obtained from the spectral action, using additional
scale for the time direction. The first equation of motion will follow from variation of 
the density with respect to this auxiliary factor $b$.  
\begin{equation}
{\cal L}  \sim  \Lambda b a^3 + 6 \dot{a}^2 a / b,
\end{equation}
giving 
\begin{equation}
6 a \dot{a}^2   -  \Lambda a^3 = 0, 
\end{equation}
and then varying $a(t)$ we obtain, 
\begin{equation}
6 \frac{d}{dt} \left(2 a \dot{a} \right) 
- 6 \dot{a}^2 - 3 \Lambda a^2 = 0,
\end{equation}
which finally gives
\begin{equation}
 12 \ddot{a} a  + 6 \dot{a}^2  - 3 \Lambda a^2 = 0.
\end{equation}
The resulting equations then read:
\begin{equation}
\frac{\dot{a}^2}{a^2}  = \frac{1}{6} \Lambda,   \qquad\qquad
\frac{\ddot{a}}{a}  = \frac{1}{6} \Lambda,  
\end{equation}
and are typical for the dark-energy dominated universe equations. 

The standard solution of the empty universe (bar the cosmological constant) 
is the exponentially growing {\bf de Sitter universe} with constant Hubble 
parameter,  
\begin{equation}
a(t) = a_0 \, exp \left( \sqrt{\frac{\Lambda}{6}} t \right). \label{empty}
\end{equation}

\subsection{An almost commutative perturbation of de Sitter universe}

In the noncommutative model, with $a_1(t)$ and $a_2(t)$ we assume 
that the effective cosmological constant is the same for both parallel
geometries and concentrate on the modification for the equations that
arise from the potential term. Using a similar procedure as in the 
nondeformed case, we introduce an auxiliary time scale $b$ (which we take
to be identical for both copies of spacetime geometry), then the potential
term scales, 
$$ b \alpha { 
(a_1 - a_2)^2 (a_2^3 +2 a_2^2 a_1 +2 a_1^2 a_2  + a_1^3)
 \over  (a_2 + a_1)^2}. $$

The set of equations of motion that arises for the full action, that involves 
the term mixing $a_1(t)$ and $a_2(t)$ is as follows:

\begin{equation}
\begin{aligned}
&6a (\dot{a}_1^2 + \dot{a}_2^2) - \Lambda (a_1^3 +a_2^3) 
- \alpha  {(a_1 - a_2)^2 (a_2^3 +2 a_2^2 a_1 +2 a_1^2 a_2  + a_1^3)
	\over  (a_2 + a_1)^2} = 0, \\
&12 \ddot{a}_1 a_1 + 6 \dot{a}_1^2 - 3 \Lambda a_1^2 
-   \alpha {(a_1 - a_2) (2 a_2^3 +2 a_2^2 a_1 +5 a_1^2 a_2  + 3 a_1^3)
	\over  (a_2 + a_1)^2} =0, \\
&12 \ddot{a}_2 a_2 + 6 \dot{a}_2^2 - 3 \Lambda a_2^2 
- \alpha  {(a_2 - a_1) (3 a_2^3 +5 a_2^2 a_1 +2 a_1^2 a_2  + 2 a_1^3)
	\over  (a_2 + a_1)^2} =0.
\end{aligned}
\end{equation}

We shall look for the perturbative solutions of the form:
$$ a_1(t) = a(t) + \epsilon r(t), \;\;\;\; a_2(t) = a(t) - \epsilon r(t),$$
bearing in mind that the above assumption might be too restrictive.

Of course, the function $a(t)$ must be the standard de Sitter, solution,
whereas for the perturbative correction we obtain in the first order 
in $\epsilon$:
\begin{equation}
12 \ddot{a}(t) r(t) +12 \ddot{r}(t) a(t) + 12 \dot{a}(t) \dot{r}(t) - 6 \Lambda a(t) r(t) 
- 6 \alpha a(t) r(t)= 0. \label{corr}
\end{equation}

\subsection{Models and solutions}

We shall consider three models to study the qualitative and significant effect of
the assumed form of the interactions. We assume that the matter or radiation terms, 
whenever occuring, are identical for both sheets, thus the equation for 
difference of the scaling factors $a_1(t)-a_2(t)$ depends only on the background
solution (which is for the solution for the identical factors $a_1(t)=a_2(t))$ and
the potential that depends on $a_1(t)$ and $a_2(t)$. 

\subsection{The empty universe}

We begin with the model of an empty universe, with the core solution (\ref{empty}).
The equation (\ref{corr}) then becomes:
\begin{equation}
6 \ddot{r}(t) +  \sqrt{6\Lambda}  \dot{r}(t) - ( 2 \Lambda + 3\alpha) r(t) = 0,
\end{equation}
and the most general solutions are:
\begin{equation}   
   c_1  e^{- \sqrt{\frac{\Lambda}{24}} + \frac{1}{4} \sqrt{6 \Lambda + 8 \alpha}}
+ c_2 e^{- \sqrt{\frac{\Lambda}{24}} - \frac{1}{4} \sqrt{6 \Lambda + 8 \alpha}}.
\end{equation} 

First of all, observe that if $\Lambda > 0$ then the solution, which shall be
of correction type will grow exponentially and is, in fact, of the same type as 
the base solution of the expanding universe. However, we need to take into
considerations the fact that $\Lambda$ is the effective cosmological constant
that was obtained from the ,,bare'' cosmological constant $\lambda$ (that came
from the heat trace expansion scaling) and the interaction terms $-\lambda^2 c$
(see in Eq. (10)). Therefore we need to consider two situations. First , if $\Lambda=0$
then core equation give the linearly growing universe, while only the corrections
give an exponential growth. From the physical point of view that is rather dissatisfying
as we can expect that the correction term $r(t)$ rather stays small when compared
to the standard evolution $a(t)$.

Another possibility is that $\Lambda < 0$, which then,  possibly reverses the roles of the ,,base'' e
 fvolution and the correction. Indeed, then the solution $a(t)$ is oscillating, whereas the correction term might add and exponential behavior 
provided that $6\Lambda + 8 \alpha >0$. So, the entire solution will, at least for 
the part of time resemble an exponential growth with a sinusoidal correction but
cannot be stable as the correction term grows too big when compared to $a(t)$.

\subsection{Radiation dominated universe}

We assume here the standard solution of a universe, in which radiation
dominates, which might have typical for the very early age evolution. We
take:
\begin{equation}
a(t) = a_0 \, \sqrt{t}, \label{radiation}
\end{equation}
which leads to the equation:
\begin{equation}
4 \ddot{r}(t) t^2 +  2 \dot{r}(t) t - (1+ 2 \Lambda t^2 + 2 \alpha t^2) r(t) = 0,
\end{equation}

The solutions are then Bessel function scaled by a time factor,
$$ r(t) \sim t^{\frac{1}{4}} J_{\frac{\sqrt{5}}{4}}(\sqrt{-2(\Lambda +\alpha)} t ), $$
which will make sense again, for $\Lambda < 0$ and $\Lambda + \alpha < 0$. 
As the Bessel functions decrease like $t^{-\frac{1}{2}}$ the correction term will also
be slowly decreasing with time. 

\subsection{Matter dominated universe}
As a last case let us see the type of corrections we might get in the case of the
standard solutions for the matter dominated universe. We have,
\begin{equation}
a(t) = a_0 \, t^{\frac{2}{3}}. \label{matter}
\end{equation}
which leads to the equation
\begin{equation}
18 \ddot{r}(t) t^2 +  12 \dot{r}(t) t - (4+ 9 \Lambda t^2 +9  \alpha t^2) r(t) = 0,
\end{equation}
and solutions
$$ r(t) \sim t^{-\frac{1}{3}} \sin \left( \frac{1}{2} \sqrt{- 2(\Lambda + \alpha)} t \right). $$
Here again, the solutions are oscillating only if $\Lambda$ and $\alpha$ are 
satisfying the same bounds as in the radiation-dominated case.

\section{Conslusions and outlook}

In the models presented above we wanted to obtain only a qualitative picture,
without discussing the values of the parameters. We have also restricted ourselves 
to very fundamental models and approximate solutions leaving the detailed
analysis of the full considered model to future work.

Nevertheless even such simplified version shows that, from point of view
of cosmology and noncommutative model-building, the scenario with cosmic
scale factor which are different for the two sheets of the two-sheeted space 
(which then bears the interpretation as the world for right-handed and 
left-handed particles) cannot be neglected. 

Though it still remains to be studied how such different cosmic scales can be 
potentially observed and, one needs to notice a lot of similarities of the above
model to the bimetric theory of gravity. It is remarkable, that 
a simple noncommutative model quite surprisingly leads to very similar Lagrangian
as an alternative theory of gravity that is considered seriously as a potential
model for the accelerating universe. The cosmological solutions of bigravity 
have been shown to reproduce the current cosmic acceleration and fitted such 
to observational data \cite{SSEMH12}. Several other papers constrained parameters
of bigravity and found that bigravity allows models that provide late-time acceleration 
in agreement with observations (for example \cite{KPA14,YHKSS15}).

It is also worth mentioning that some general noncommutative models, with deformed
space-time effectively lead to a version of action and metric fields that in the
classical limit reduce themselves to a bimetric gravity models \cite{CSV18}. 

The presented model needs to be extended to the full version of Connes' Standard Model
\cite{CC12} with a full algebra and the resulting terms of the spectral action (even beyond the
second leading term). Only then a detailed analysis of the possible values of the
parameters as well as the observational constraints can be carried out, and we plan
to proceed with the analysis in the forthcoming work.

\end{document}